\documentclass[pdflatex,sn-mathphys-num]{sn-jnl}

\usepackage{graphicx}%
\usepackage{multirow}%
\usepackage{amsmath,amssymb,amsfonts}%
\usepackage{amsthm}%
\usepackage{mathrsfs}%
\usepackage[title]{appendix}%
\usepackage{xcolor}%
\usepackage{textcomp}%
\usepackage{manyfoot}%
\usepackage{booktabs}%
\usepackage{algorithm}%
\usepackage{algorithmicx}%
\usepackage{algpseudocode}%
\usepackage{listings}%
\usepackage{caption}
\usepackage{subcaption}

\theoremstyle{thmstyleone}%
\theoremstyle{thmstyletwo}%

\theoremstyle{thmstylethree}%

\begin{document}

\title[A Method for Constructing Quasi-Random Peaked Quantum Circuits]{A Method for Constructing Quasi-Random Peaked Quantum Circuits}


\author*[1]{\fnm{Oleg} \sur{Udalov}}\email{udalovog@gmail.com}

\affil*[1]{\orgdiv{Independent Researcher}, \orgaddress{\postcode{20878}, \state{MD}, \country{USA}}}


\abstract{An algorithm is proposed for constructing quasi-random ``peaked'' quantum circuits - i.e., circuits whose final qubit state exhibits a high probability concentration on a specific computational basis state. These circuits consist of random gates arranged in a brick-wall architecture. While the multiqubit state in the middle of the circuit can exhibit significant entanglement, the final state is, with high probability, a predetermined pure bitstring. A technique is introduced to obscure the final bitstring in the structure of the quantum circuit.  The algorithm allows precise control over the probability of the final peaked state. A modified version of the algorithm enables the construction of double- or multi-peaked quantum circuits. The matrix product state (MPS) method is evaluated for simulating such circuits; it performs effectively for shallow peaked circuits but offers no significant advantage for deeper ones.}





\maketitle

\section{Introduction}\label{sec1}

Quantum advantage is a cornerstone of quantum computing development \cite{Intro_suprem_1, Intro_suprem_2, Intro_suprem_3, Intro_suprem_4, Intro_suprem_5}. Demonstrating quantum advantage requires quantum algorithms that, on one hand, provide a convincing speedup compared to classical algorithms, and on the other hand, are verifiable. While some quantum algorithms (QAs), such as Shor's algorithm~\cite{Shor}, are proven to offer substantial speedup and can be easily verified, they are mostly infeasible to run on current NISQ quantum machines. 

Random quantum circuits (RQCs) are considered a promising candidate to demonstrate and verify quantum advantage. Google claimed quantum advantage by running a random circuit that could not be simulated using a classical computer~\cite{Intro_suprem_1}. An RQC of general form generates a complex, strongly entangled state of qubits at the end. While this state is difficult (or even impossible) to simulate on a classical computer within a reasonable time, it is also challenging to verify whether the quantum computer performs the computation correctly. To address this, a specific class of random (or quasi-random) circuits was recently proposed in Ref.~\cite{Peaked_main}, known as peaked quantum circuits (PQCs).

A PQC is a quasi-random quantum algorithm whose final state is concentrated on a single bitstring. While direct simulation of such a PQC on a classical computer is difficult (or even impossible) within a reasonable time, the result of the quantum computation can be easily verified.

The PQC can be considered as an example of a quantum circuit Born machine (QCBM) ~\cite{PhysRevA.98.062324, Benedetti2019Generative, Du2022Power, Coyle2021Quantum} with a specific probability distribution heavily weighted on a single bitstring. Several ways of creating various Born machines have been studied previously. Some follow the approach where the quantum circuit is ``trained'' using a quantum-classical hybrid approach~\cite{Wecker2016Training}. In this approach, the quantum circuit has a fixed structure but tunable parameters. These parameters are optimized so that the desired probability distribution of the circuit output is achieved. More advanced techniques use, for example, hierarchical learning~\cite{Gharibyan2023Hierarchical}.

In Ref.~\cite{Peaked_main}, the PQC is created as follows. First, a random brick-wall circuit~\cite{PhysRevResearch.4.023097, Mihalikova2025Impact, PhysRevA.102.012415} is generated. Such a circuit transforms the initial state $|0\rangle^n$ into a complex qubit state with a ``uniform'' distribution of sampled bitstrings across the entire Haar space. Then, the second part of the circuit is generated in a way that brings the random state back to the initial state $|0\rangle^n$ or any other desired bitstring. In general, one can consider the second half of the circuit as the inverse of the first half plus a set of NOT gates to obtain the desired bitstring at the end. The peculiarity of this work is that they were able to use fewer gates (fewer layers of the brick-wall circuit) in the second half than in the first half. The disadvantage of the proposed method is that the creation of the second half of the circuit is not scalable. They use an optimization procedure to find the entire second half at once. Thus, as the number of qubits and circuit depth increase, the number of optimized parameters grows, and the optimization time quickly becomes impractical.

Here, we propose a scalable method to create quasi-random peaked circuits with any given final peaked bitstring. In contrast to Ref.~\cite{Peaked_main}, where the entire second half (the inverse of the first half) is generated through a single optimization process, we perform the optimization piece by piece. The size of each piece may vary. In addition, the algorithm allows the creation of PQCs with multiple peaks and provides control over the peakedness.

The paper is organized as follows. In Sec.~\ref{sec2}, the method for constructing the quasi-random peaked circuit with a hidden bitstring is described. We discuss the properties of the PQC in Sec.~\ref{sec3} and introduce the parameter controlling the peakedness of the circuit. Sec.~\ref{sec4} discusses the extension of the method to generate multipeaked random circuits. The next section proposes how the size of the random circuits can be adjusted. Finally, in Sec.~\ref{sec6}, we discuss whether the PQC created using the proposed method can be ``resolved'' (i.e., the hidden bitstring is found) using the matrix product state method.

\section{PQC creation algorithm}\label{sec2}

The algorithm to create a peaked circuit consists of three steps. In the first step, we prepare a random circuit described by an operator \(\hat{Q} = \prod_{i=1}^{N} B_i\) acting on all qubits, where \(\{B_i\}\) is an ensemble of random two-qubit operators and \(\prod_{i=1}^{N}\) denotes ordered multiplication. Then, we create the inverse of this operator as \(\hat{Q}^{-1} = \prod_{i=N}^{1} B_i^{-1} = \prod_{i=N+1}^{2N} \widetilde{B}_i\) by inverting each random element \(B_i\) of the operator \(\hat{Q}\) and reversing the order of the operators. Here, we define \(\widetilde{B}_i = B_{N + 1 - i}^{-1}\) for \(i > N\) and \(\widetilde{B}_i = B_i\) for \(i \leq N\). We call the second half of the circuit the ``mirror'' part. Combining \(\hat{Q}^{-1} \hat{Q} = \hat{I}\), we obtain a peaked circuit with the final state \(|00\ldots0\rangle\). The total circuit can be represented as \(\hat{Q}^{-1} \hat{Q} = \prod_{i=1}^{2N} \widetilde{B}_i\). This is a trivial peaked circuit which can be easily recognized by comparing elements. For this circuit, the symmetry relation \(\widetilde{B}_i = \widetilde{B}_{2N +1 - i}^{-1}\) is obeyed. The random circuits constructed in this way were extensively considered for gates errors measurements~\cite{Emerson2005Scalable, Elben2023Randomized}.

At the second step, we change the final state of the circuit from \(|00\ldots0\rangle\) to some string \(x_{\text{hid}} = |101001011\ldots10\rangle\). To do this, we apply NOT gates to the corresponding qubits at the beginning of the circuit and then move the NOT gates through the circuit using commutation relations. Moving them in a random way and merging them with the operators \(\widetilde{B}_i\) allows us to hide the bitstring (i.e., to conceal which qubits are affected by the NOT gates). Note, however, that finding the hidden bitstring is still relatively simple. One can use the symmetry relations between the first and second halves of the circuit. Comparing operators \(\widetilde{B}_i\) in the first and second halves of the circuit should allow identification of the NOT gates.

At the third stage, we modify the second (``mirror'') half of the circuit, \(\hat{Q}^{-1} \rightarrow \hat{Q}_1\), in such a way that \(\hat{Q}_1 \approx \hat{Q}^{-1}\), but the symmetry \(\widetilde{B}_i = \widetilde{B}_{2N + 1 - i}^{-1}\) is broken. Let us now consider each step in more detail. The three steps of the circuit creation are illustrated in Fig.~\ref{fig_circ_create}.

\subsection{Step 1. Creating a random quantum circuit}\label{subsec2_1}
\subsubsection{Building block}\label{subsubsec2_1_1}

\begin{figure}[h]
\centering
\includegraphics[width=1\textwidth]{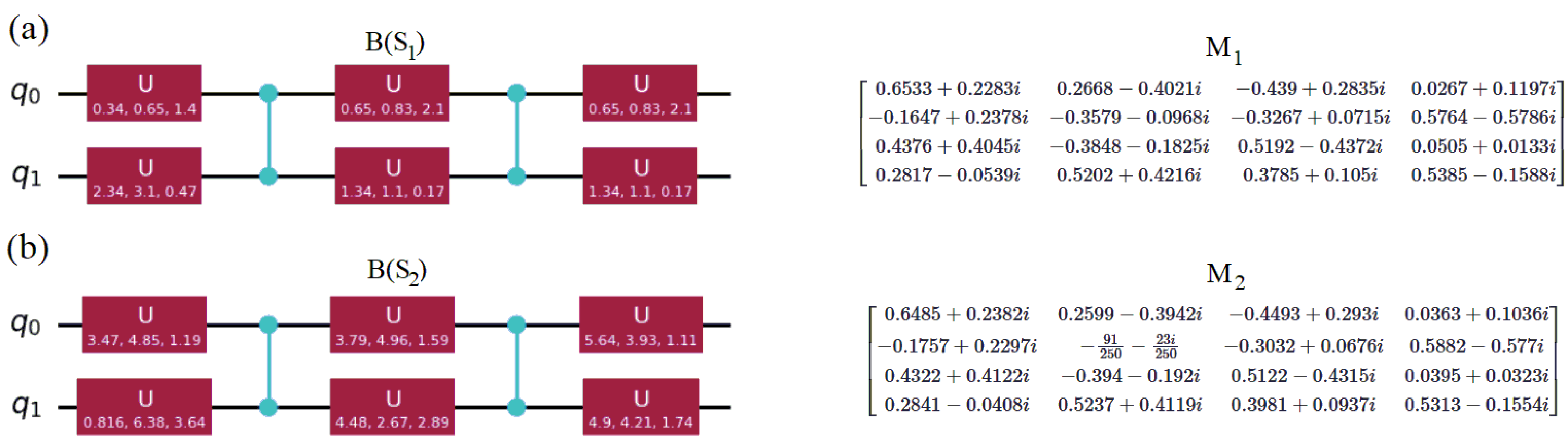}
\caption{(a) A building block of the RQC, \(B(S_1)\), and the corresponding matrix \(M_1\) describing this block. $S_1$ is the set of angles defining \(U_3(\theta, \phi, \lambda)\) gates.  (b) An alternative block \(B(S_2)\), which has a completely different set of parameters but produces a similar matrix, \(M_2 \approx M_1\). The figure is generated using Qiskit~\cite{JavadiAbhari2024Qiskit}.}\label{fig1}
\end{figure}

Here, we construct a random quantum circuit with \(n_\text{q}\) qubits. A building block \(B_i\) of the circuit is a two-qubit operator consisting of six \(U_3(\theta, \phi, \lambda)\) gates and two CZ gates. This operator was referred to as \(B_i\) in the previous section. It is shown in Fig.~\ref{fig1}(a).

The block \(B\) is defined by 18 rotational angles (parameters \(S\)) and represents a two-qubit operator, or equivalently, a matrix \(M\) of size \(4 \times 4\). Importantly, blocks with the same (or nearly the same) matrix can be generated by multiple different sets of angles. That is, for any block \(B(S_1)\) with matrix \(M_1\), there exists at least one block \(B(S_2)\) with matrix \(M_2\) such that \(M_1 \approx M_2\). An example of such a pair of blocks is shown in Fig.~\ref{fig1}. 

The difference between the blocks can be calculated as follows

\begin{equation}
\tilde \delta=\sqrt{\sum_{ij}\left({M}_1^{ij}-{M}_2^{ij}\right)^2/16}, \label{eq_delta_i}
\end{equation}
where $M_{1,2}^{ij}$ are the matrix elements of the \(4 \times 4\) matrices describing blocks $B(S_1)$ and $B(S_2)$.

Another important property of the block is that it can be commuted with a NOT gate. If a NOT gate is applied before the block, it can be moved through the block to the opposite side, resulting in a modified version of the block.

\subsubsection{Combining blocks into a RQC}\label{subsubsec2_1_2}
\begin{figure}[h]
\centering
\includegraphics[width=1\textwidth]{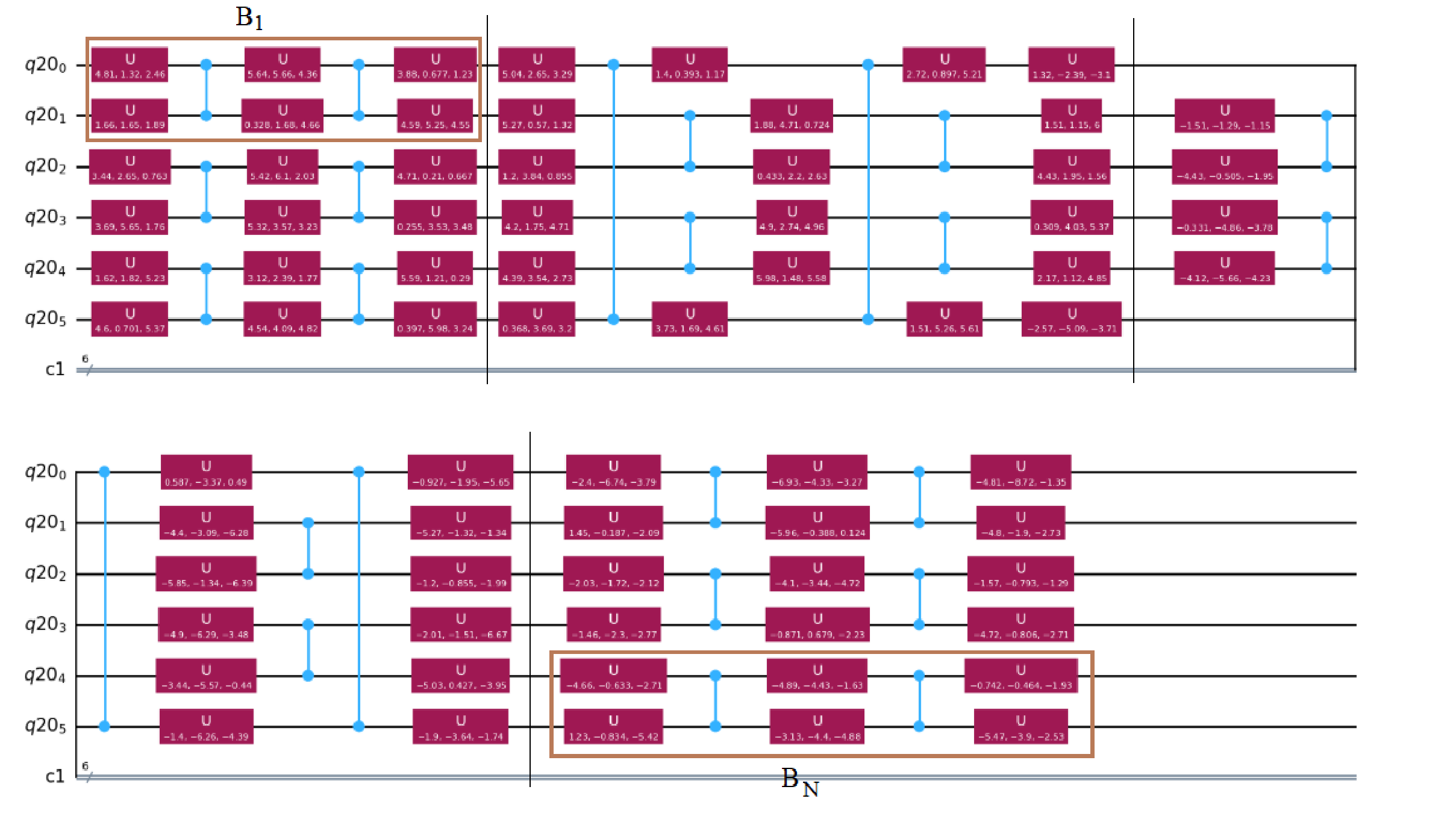}
\caption{RQC with \(n_\text{l} = 4\) layers and \(n_\text{q} = 6\) qubits. The total number of blocks is \(N = 12\). The first and last blocks are highlighted with brown rectangles. Black lines separate the circuit layers.}\label{fig2}
\end{figure}

The circuit resembles a brick-wall structure and acts on an even number of qubits, \(n_\text{q}\). One layer of the circuit contains \(\frac{n_\text{q}}{2}\) blocks, each applied to a different pair of qubits. In total, the circuit consists of \(n_\text{l}\) layers, so the total number of blocks is \(N = \frac{n_\text{q} n_\text{l}}{2}\). In the odd-numbered layers, the blocks are applied to the following pairs of qubits: \((1,2), (3,4), (5,6), \ldots\). In the even-numbered layers, the block positions are shifted: \((n_\text{q}, 1), (2,3), (4,5), \ldots\). Note that the parameter sets \(S_i\) for each block are generated randomly. This results in a brick-wall-type quantum circuit. All the blocks in this circuit collectively form the multi-qubit operator \(\hat{Q}\). An example of such a circuit is shown in Fig.~\ref{fig2}.

\begin{figure}[h]
\centering
\includegraphics[width=0.8\textwidth]{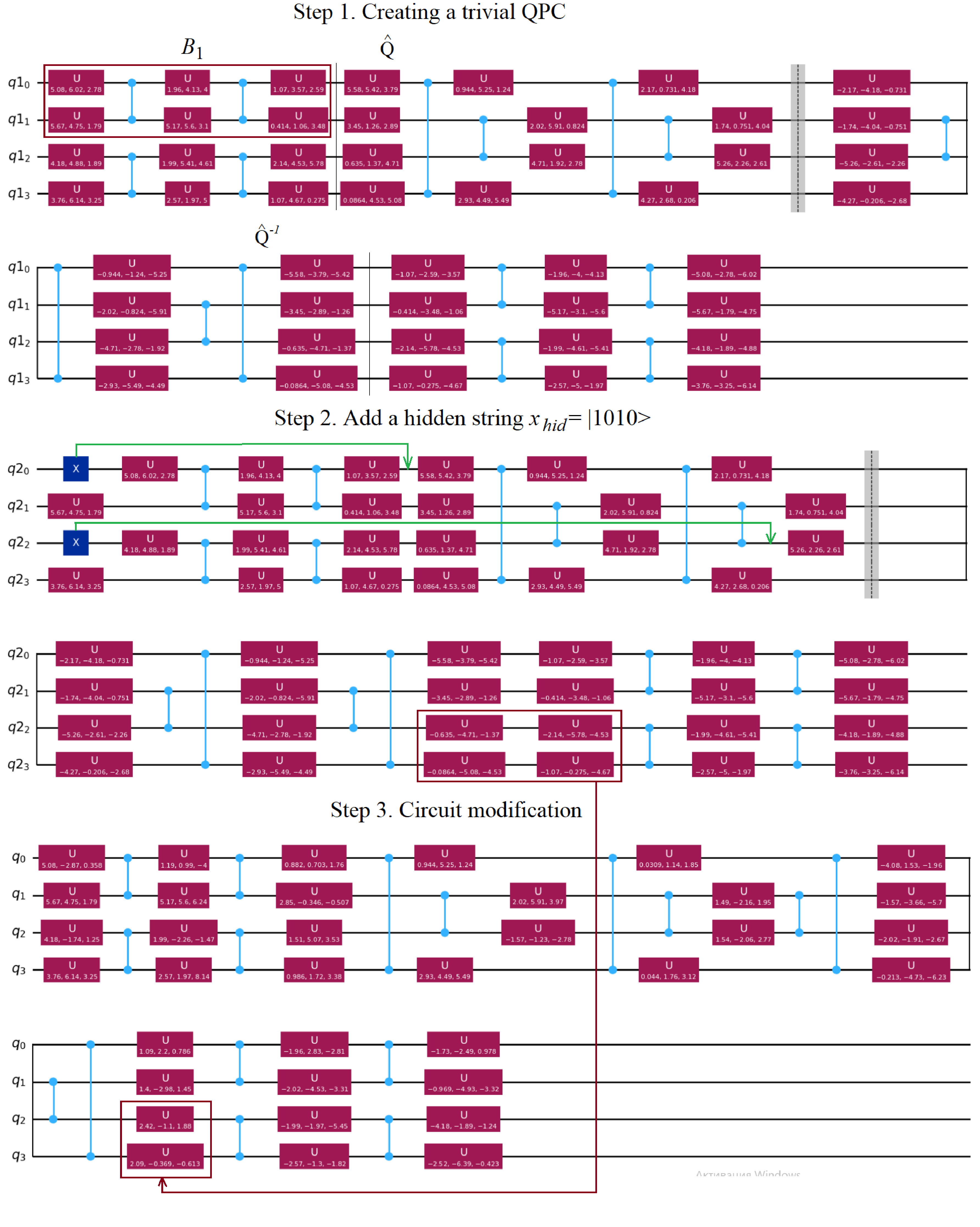}
\caption{Three steps to create a quasi-random PQC. Step 1: a 4-qubit circuit is constructed with \(n_\text{l} = 2\) layers. The random circuit \(\hat{Q}\) is divided into layers by solid black vertical lines. A dashed vertical line separates \(\hat{Q}\) and \(\hat{Q}^{-1}\). Step 2: a hidden bitstring \(|1010\rangle\) is embedded. Green arrows indicate where the corresponding NOT gates are moved using the commutation relations given in Eqs.~(\ref{eq1} - \ref{eq3}). Step 3: Eq.~(\ref{Approx_trans}) is used to modify each block in the second half of the circuit. Double \(U_3\) blocks are then merged.}\label{fig_circ_create}
\end{figure}

\subsubsection{Creating symmetric (``mirror'') second half of the circuit}\label{subsubsec2_1_3}
Inversion of a block \(B_i\) is straightforward. To obtain the matrix \(\hat{Q}^{-1}\), we apply the blocks \(B_i^{-1}\) in the reverse order. Thus, the entire circuit satisfies \(\hat{Q} \hat{Q}^{-1} = \hat{I}\), where \(\hat{I}\) is the identity operator. The identity matrix obviously corresponds to a peaked circuit. An example of this ``ideal'' peaked circuit is shown in Fig.~\ref{fig_circ_create} in Step 1.

\subsection{Step 2. Embedding NOT gates.}\label{subsec2_2}
At the second step of the circuit creation (see Fig.~\ref{fig_circ_create}), we choose the hidden bitstring \(x_{\text{hid}}\). According to this string, we insert NOT gates at the beginning of the circuit. The next step is to move these NOT gates through the circuit to randomized positions. To move the gates through the circuit, we use the following relations:
\begin{equation}
X\ U_3\left(\theta,\phi,\lambda\right)=U_3\left(\theta,\pi-\phi,\pi-\lambda\right)X\exp{(i(\phi+\lambda))}, \label{eq1}
\end{equation}
\begin{equation}
U_3\left(\theta,\phi,\ \lambda\right)X=U_3\left(\pi-\ \theta,\ -\phi,\lambda-\pi\right)\exp{(i\phi)},\label{eq2}
\end{equation}
\begin{equation}
X\left(2\right)CZ\left(1,2\right)=CZ(1,2)Z(1)X(2).\label{eq3}
\end{equation}

Using these relations, one can move the NOT gates to random positions within the circuit without modifying its overall structure (brick-wall architecture). Adding and moving the NOT gates partially breaks the circuit's symmetry. However, it remains relatively easy to identify the location of each NOT gate. By checking the equality \(\widetilde{B}_i = \widetilde{B}_{2N + 1 - i}^{-1}\), one can find all blocks that were not modified and, consequently, recover the hidden string. Therefore, further modification of the circuit is necessary.

\subsection{Step 3. Circuit modification.}\label{subsec2_3}

The symmetry of the circuit allows one to find the hidden string by comparing blocks \(i\) and \(2N + 1 - i\). Thus, recovering the hidden string reduces to comparing block pairs, which is not a difficult mathematical problem. To increase the difficulty, we apply two types of modifications. As mentioned before, the block \(\widetilde{B}_i\) is not uniquely defined by its 18 parameters; almost the same block can be realized with different sets of rotational angles. The first modification is as follows: for each block \(\widetilde{B}_i\) with parameters \(S_i\) and matrix \(M_i\) in the ``mirror'' half of the circuit (\(i > N\)), we find a block \(\widetilde{B}_i^{\text{mod}}\) with a significantly different parameter set \({S}_i^{\text{mod}}\) and matrix \(M_i^{\text{mod}}\) such that

\begin{equation}
M_i^{\text{mod}}\approx M_i. \label{Approx_trans}
\end{equation} 

The approximate transformation can be performed using an optimization procedure as follows. Given a matrix \(M_i\) corresponding to the block \(\widetilde B_i\) and a fixed block structure defined by 18 angles, one searches the 18-dimensional parameter space for a block with matrix \(M_i^{\text{mod}} \approx M_i\). A gradient descent method (or any other optimization technique) can be employed to find a set of angles that satisfy this condition. We used the ``COBYLA'' optimizer for this purpose. A random starting point in the optimization ensures that the resulting parameter set \({S}_i^{\text{mod}}\) differs substantially from the initial set \(S_i\).

In the previous section, we introduced the difference (deviation) between the initial and alternative blocks, denoted by \(\tilde \delta_i\), Eq.~(\ref{eq_delta_i}). Here, we define the average deviation of the modified circuit as the mean value of \(\tilde \delta_i\) taken over all modified blocks

\begin{equation}
\delta =\sum_{N+1}^{2N}\tilde\delta_i/N.
\end{equation}

The summation is taken over the second (``mirror'') half of the circuit, where the modifications are applied.
Now, direct comparison of block parameters no longer suffices to compare the blocks. However, the matrices describing the blocks can still be compared. The second step is to join the blocks: each block ends with two \(U_3\) gates and begins with two \(U_3\) gates, which can be combined. Once the blocks are merged, the comparison of \(\widetilde B_i\) and \(\widetilde B_{2N + 1 - i}\) cannot be performed individually for each pair. Instead, comparing \(\widetilde B_i\) and \(\widetilde B_{2N - i}\) requires simultaneous comparison of blocks \(\widetilde B_{i-1}\) versus \(\widetilde B_{2N - i + 2}\) and blocks \(\widetilde B_{i+1}\) versus \(\widetilde B_{2N - i}\). This necessitates comparing all blocks simultaneously rather than pairwise. Such a task is mathematically much more challenging, and the computational complexity scales rapidly with the circuit size. We now have \(N/2\) coupled nonlinear equations over \(18 \times N/2\) variables. The final RPC is shown in Step 3 of Fig.~\ref{fig_circ_create}.

\section{Properties of a quasi-random PQC}\label{sec3}
The method to create the random PQC uses the approximate transformation given by Eq.~(\ref{Approx_trans}). Due to the approximate nature of this transformation, the final circuit is not an ideal peaked circuit. This implies that the final state of the circuit (even in the absence of any gate execution errors) is not exactly the hidden bitstring \(x_{\text{hid}}\), unlike the initial random circuit before the transformation.

One can introduce measures to characterize how peaked the circuit is. The simplest way to estimate this is by comparing the probabilities of different final states of the circuit. A peaked circuit assumes that the probability \(P_{\text{peak}} = P(x_{\text{hid}})\) is much higher than the probability of any other state. In a completely random circuit, one expects a uniform distribution of probabilities over all bitstrings, with an average value of \(1/2^{n_\text{q}}\). In a peaked circuit, we have \(P_{\text{peak}} \gg 1/2^{n_\text{q}}\). In the proposed algorithm, the probability \(P_{\text{peak}}\) depends on the average block deviation \(\delta\). If \(\delta = 0\), the circuit should be ideally peaked with \(P_{\text{peak}} = 1\). With increasing $\delta$ the peakedness ($P_\text{peak}$) goes down. We create and simulate multiple random peaked circuits using Qiskit~\cite{JavadiAbhari2024Qiskit} to study how the probability \(P_{\text{peak}}\) depends on the circuit size and the average deviation \(\delta\).

\begin{figure}[h]
\centering
\includegraphics[width=1\textwidth]{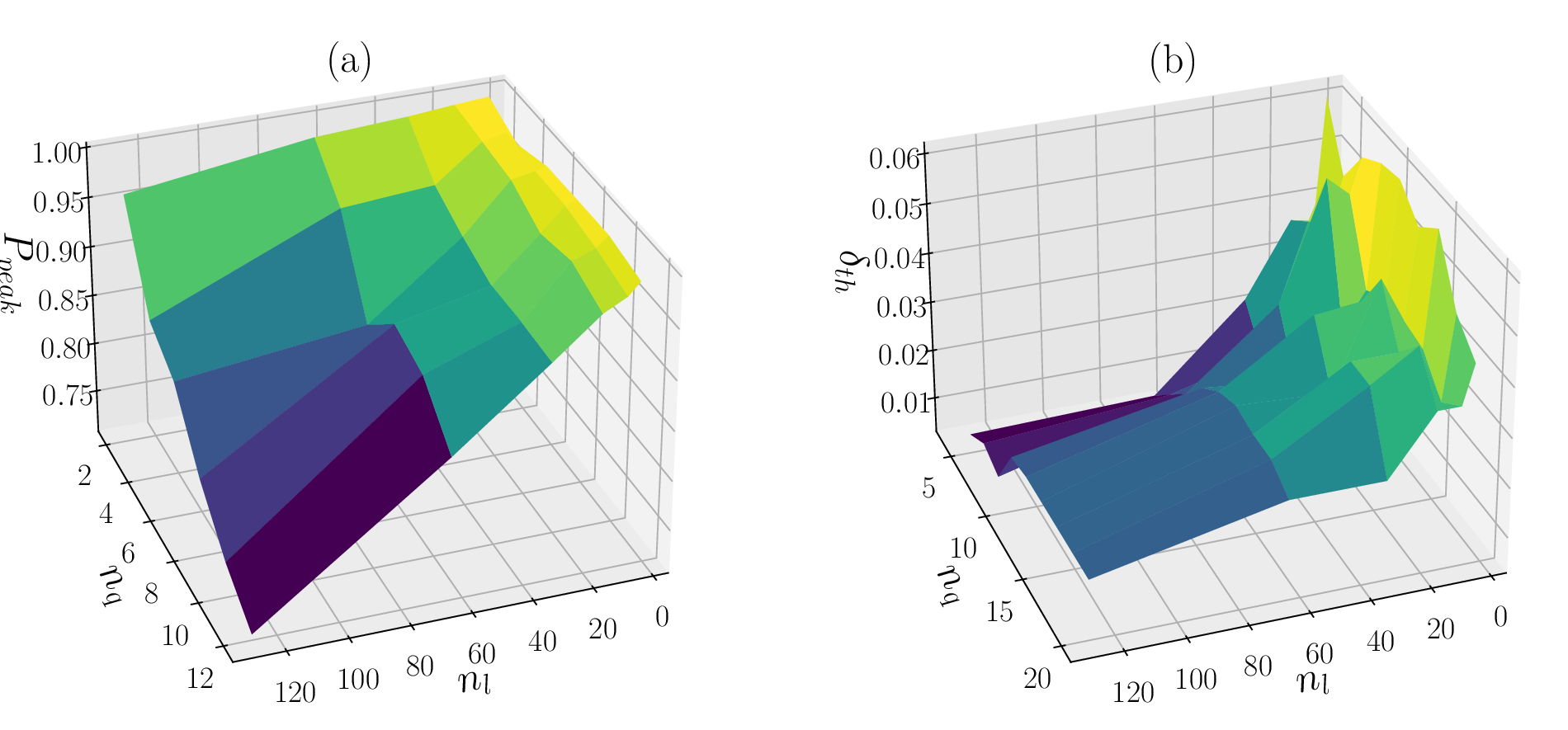}
\caption{(a) Peakedness (probability to observe the hidden string \(x_{\text{hid}}\), \(P_{\text{peak}}\)) as a function of circuit size, characterized by the number of qubits \(n_\text{q}\) and the number of layers \(n_\text{l}\).  
(b) Maximum allowed deviation \(\delta_{\text{th}}\) for which the circuit remains peaked with \(P_{\text{peak}} / P_{\text{second}} = 10\), plotted as a function of \(n_\text{q}\) and \(n_\text{l}\).}\label{fig_prob_1}
\end{figure}

Figure~\ref{fig_prob_1}(a) demonstrates how the peakedness, measured by the peak probability \(P_{\text{peak}}\), depends on the circuit size (number of qubits and circuit depth) for a given block deviation \(\delta\). The chosen block deviation is approximately 0.3\% (\(\delta = 0.003\)). One can see that the peakedness \(P_{\text{peak}}\) decreases approximately linearly with both the number of qubits and the number of layers in the circuit. Based on the figure, one can estimate how quickly \(P_{\text{peak}}\) approaches zero as the circuit size increases. For example, for 60 layers, a linear interpolation suggests that \(P_{\text{peak}}\) would vanish around 40 qubits. 

Note, however, that this estimate is not entirely accurate. By definition, a peaked circuit requires that the probability of the hidden string is significantly higher than that of other states. To quantify this, we introduce the ratio between the probability of the peak state \(x_{\text{hid}}\) and the probability of the second most probable state, \(P_{\text{second}}\). If this ratio, \(P_{\text{peak}} / P_{\text{second}}\), is sufficiently large, the circuit can be considered peaked. In this work, we arbitrarily choose a critical threshold of \(P_{\text{peak}} / P_{\text{second}} = 10\) to classify circuits as peaked.

Figure~\ref{fig_prob_1}(b) shows the maximum allowable deviation \(\delta_\text{th}\) required to obtain a peaked circuit (\(P_{\text{peak}} / P_{\text{second}} = 10\)) as a function of the number of qubits \(n_\text{q}\) and the number of layers \(n_\text{l}\) (in the first half of the circuit). It can be observed that this maximum deviation decreases with increasing circuit depth. However, it depends only weakly on the number of qubits. Notably, for \(n_\text{l} = 128\), the maximum deviation is approximately 0.015 across all qubit numbers, except for \(n_\text{q} = 4\) and \(6\), for which the behavior is unclear.

Based on this data, one can expect that the proposed method may be effective for large quantum circuits  - beyond the reach of classical simulation - with several tens of qubits.

Another conclusion is that the peakedness can be effectively controlled through the parameter \(\delta\).

\begin{figure}[ht]
    \centering
    \begin{subfigure}{1\textwidth}
    	\caption{Adding entangling block}
        \includegraphics[width=\linewidth]{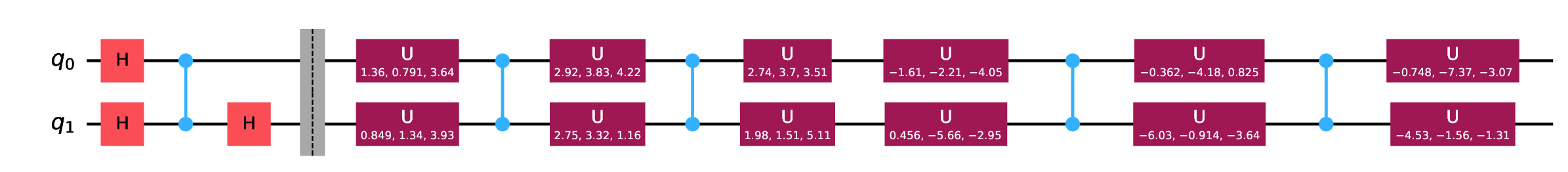}
        
    \end{subfigure}
    \hfill
    \begin{subfigure}{1\textwidth}
    	\caption{Absorbing entangling block to the next block}
        \includegraphics[width=\linewidth]{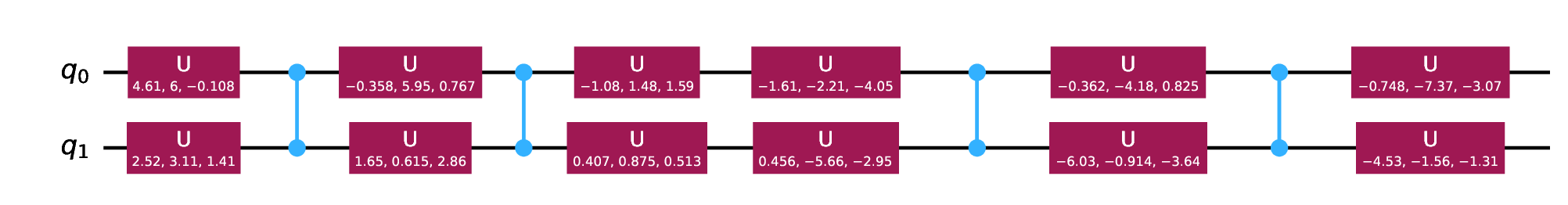}
        
    \end{subfigure}
    \caption{Creating a quasi-random peaked circuit with two peaks, \(|00\rangle\) and \(|11\rangle\). In the first step (a), an entangling block is added at the beginning of the circuit. In the second step (b), this entangling block is ``absorbed'' into the circuit. In general, the entangling block can be moved to an arbitrary position within the circuit.}\label{fig_double_peak}
\end{figure}

\section{Creating a double peaked circuit with two hidden strings}\label{sec4}
The algorithm described in Sec.~\ref{sec2} creates quasi-random peaked circuits with a probability peak on a single measurement basis string. It can be straightforwardly generalized to generate double-peaked (and, more generally, multi-peaked) probability distributions. For example, one can add entangling gates at the beginning of the circuit, as shown in Fig.~\ref{fig_double_peak}. Such a gate consists of single-qubit rotations (Hadamard gates) and a CZ gate. This block can be easily absorbed into the building blocks of our circuit, as illustrated in Fig.~\ref{fig_double_peak}(b). Furthermore, it can be moved through the circuit to a random position, allowing it to be hidden. The entanglement block in Fig.~\ref{fig_double_peak} creates two peaks at the strings \(|00\rangle\) and \(|11\rangle\). By using NOT gates and more complex entangling operators, one can generate any desired pair of strings.

\begin{figure}[ht]
    \centering
    \begin{subfigure}{1\textwidth}
    	\caption{4-qubit block}
        \includegraphics[width=\linewidth]{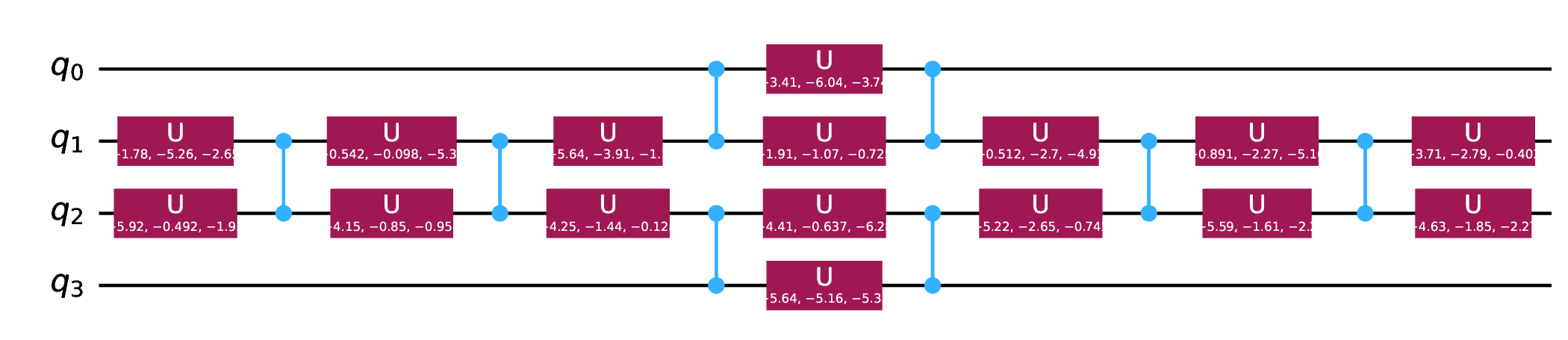}
        
    \end{subfigure}
    \hfill
    \begin{subfigure}{1\textwidth}
    	\caption{Approximately equivalent 4-qubit block of smaller size}
        \includegraphics[width=\linewidth]{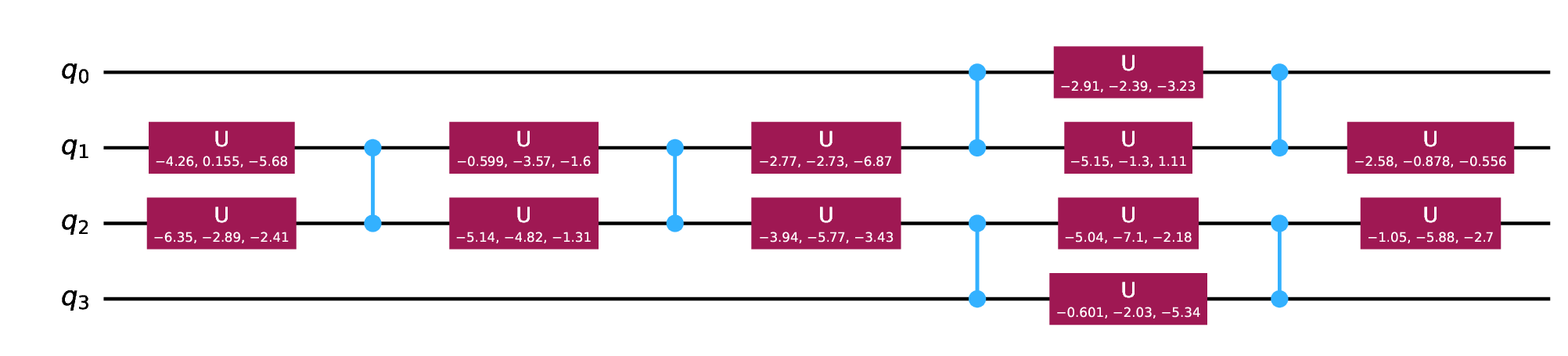}
        
    \end{subfigure}
    \caption{Reducing the block size. (a) Initial block, $B_4$.(b) Block with a reduced size $B_4^\text{red}$. The parameters of the reduced block can be obtained via an optimization procedure or using machine learning techniques.}\label{fig_block_reduction}
\end{figure}

\section{Changing the size of the ``mirror'' part of the circuit (second half of the circuit)}\label{sec5}
One of the findings in Ref.~\cite{Peaked_main} is that the second half of the circuit (which inverts the random first part) can be made shorter than the first random part by using fewer layers. To achieve this, the authors choose a shorter second part and then optimize all the angles of the \(U_3\) gates to approximate the inverse of the operator \(\hat{Q}\). In contrast, we propose not to optimize all the angles for a fixed size of the second half. Instead, we gradually shrink the size of the second half, starting from the exact inverse matrix (or from the modified inverse matrix obtained in Sec.~\ref{subsec2_3}).

As in the previous section, we replace certain blocks of the circuit with blocks that perform approximately the same function but have a shorter size. In particular, one can demonstrate that the 4-qubit block shown in Fig.~\ref{fig_block_reduction}(a) (denoted as \(B_4\)) can be replaced by the block shown in Fig.~\ref{fig_block_reduction}(b) (denoted as \(B_4^{\text{red}}\)). Finding the approximate block \(B_4^{\text{red}}\) can be achieved via an optimization procedure. First, the matrix corresponding to the original 4-qubit block \(B_4\) is computed. Then, the angles of the \(U_3\)-gates in the reduced block \(B_4^{\text{red}}\) are optimized to approximate the same operation. Notably, the smaller the rotation angles in the original block \(B_4\), the higher the precision achievable in approximating \(B_4^{\text{red}}\). After the optimization, one gets $M_4^{\text{red}}\approx M_4$, where $M_4$ and $M_4^{\text{red}}$ are the matrices corresponding to the blocks.

\begin{figure}[h]
\centering
\includegraphics[width=1\textwidth]{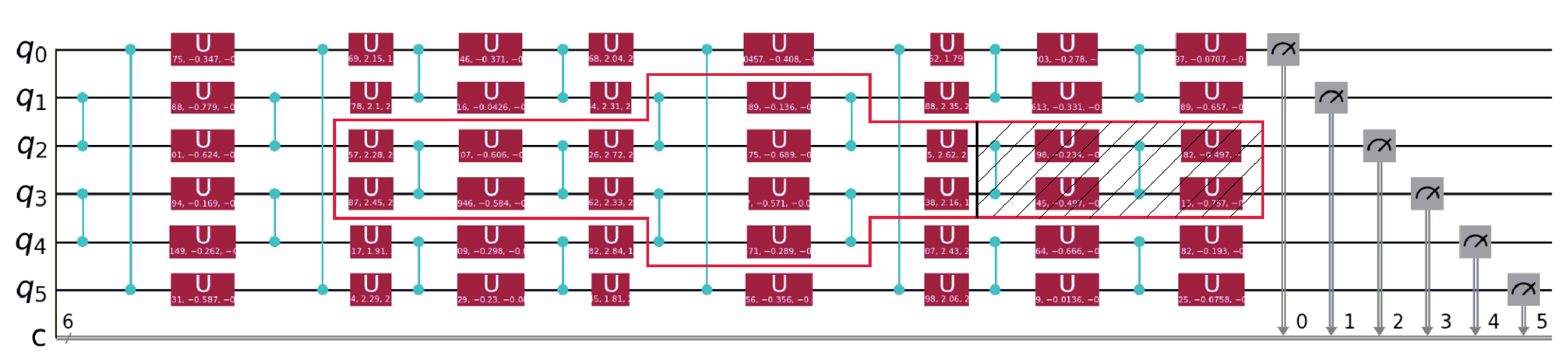}
\caption{Last portion of the 6-qubit PQC. The region enclosed by the red line corresponds to the block $B_4$ shown in Fig.~\ref{fig_block_reduction}(a). The shaded area indicates the portion that would be removed after replacing it with the shorter block shown $B_4^\text{red}$. Other blocks at the end of the circuit can be similarly replaced, allowing the entire last layer to be removed.}\label{Fig_shrinking}
\end{figure}

To shrink the circuit size, we need to identify blocks \(B_4\) at the end of the brick-wall circuit, as shown in Fig.~\ref{Fig_shrinking}. All blocks at the edge should be replaced with shorter blocks \(B_4^{\text{red}}\). Since these blocks are located at the end of the peaked circuit, the final state of the qubits after each block is well defined by the hidden string \(x_{\text{hid}}\). Therefore, only certain elements of the matrix representing the block \(B_4^{\text{red}}\) need to match (or closely approximate) the corresponding elements in the matrix of \(B_4\). For example, if the final state of the peaked circuit is \(|0\rangle^{n}\), then only the first row of the matrices for \(B_4\) and \(B_4^{\text{red}}\) need to coincide. This feature allows for improved precision in the approximation.

After all blocks at the circuit edge are replaced, the circuit becomes one layer shorter. Due to the approximate nature of the block replacement, the probability of the peak bitstring decreases. Depending on the initial peakedness and the precision of the replacement, this method allows cutting few layers. 

\begin{figure}[h]
\centering
\includegraphics[width=1\textwidth]{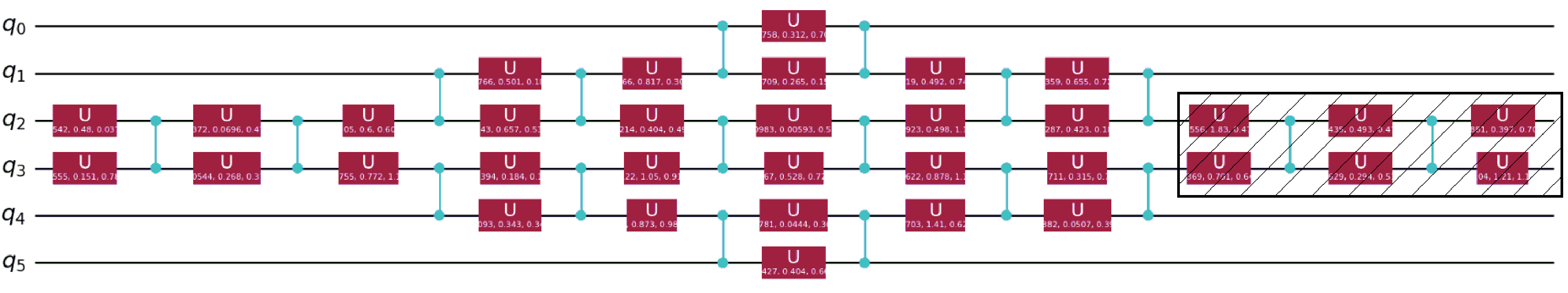}
\caption{A 6-qubit block that can be modified to shrink a circuit. The shaded region represents the portion that can be removed (with additional modification of the rest of the $U_3$ gates), while the matrix describing the entire block remains nearly identical to that of the original block.}\label{Fig_shrinking_block_6}
\end{figure}

One way to increase the precision of the replacement is to enlarge the size of the block being replaced. For example, see Fig.~\ref{Fig_shrinking_block_6}. Blocks of other shapes can also be employed.

Note that this circuit reduction procedure can also aid in ``resolving'' the peaked circuit. If the block replacement is sufficiently precise, the circuit depth can be reduced while maintaining a high level of peakedness.

\section{Using matrix product state to find the hidden string}\label{sec6}
As the number of qubits increases (beyond approximately 30~\cite{Xu2023Herculean}), the complexity of direct classical simulation of quantum circuits surpasses the capabilities of classical computers. However, there exist approximate simulation methods that enable simulation of much larger circuits. One such method is the matrix product state (MPS) approach~\cite{PhysRevA.73.022344, PerezGarcia2007MPS, Martin2024, PhysRevResearch.6.013326}. In this method, multiqubit states with low entanglement can be simulated efficiently even for large numbers of qubits. The amount of entanglement accounted for is characterized by the so-called bond dimension \(\chi\). The lowest amount of entanglement is captured when \(\chi = 1\), whereas the highest corresponds to \(\chi = 2^{n_\text{q}/2}\). Selecting a low \(\chi\) significantly speeds up simulations and allows for simulating quantum circuits with many qubits.

While the peaked circuit produces a final state that is a pure, unentangled state in the measurement basis, the intermediate states within the circuit can be strongly entangled. Therefore, depending on the amount of entanglement, the peaked circuit may or may not be efficiently simulable using the MPS method.

\begin{figure}[h]
\centering
\includegraphics[width=1\textwidth]{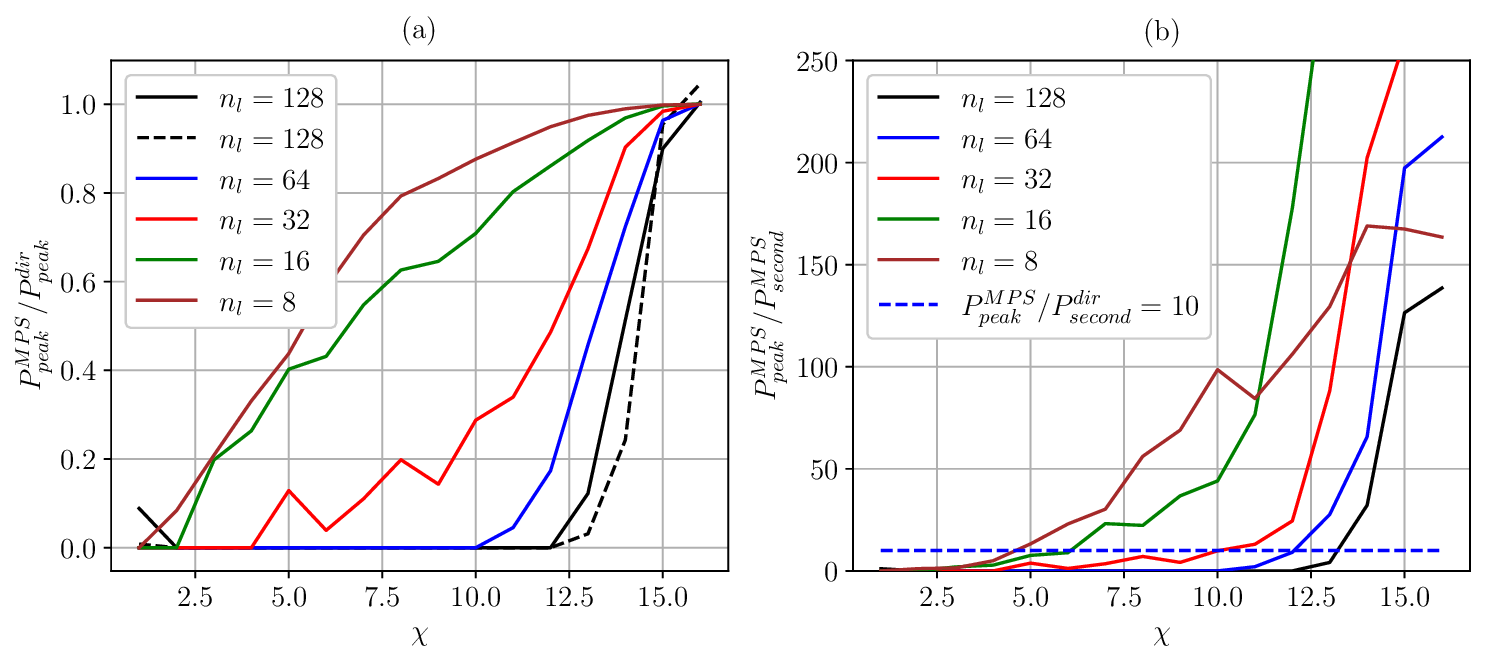}
\caption{Relative probabilities of observing the hidden string at the circuit output as a function of the bond dimension \(\chi\).  
The considered circuit has 8 qubits.  
(a) The solid line shows the ratio \(P_{\text{peak}}^{\text{MPS}} / P_{\text{peak}}^{\text{dir}}\) for different numbers of circuit layers \(n_\text{l}\).  
The dashed line shows the normalized ratio \(\left(P_{\text{peak}}^{\text{MPS}} / P_{\text{second}}^{\text{MPS}}\right) / \left(P_{\text{peak}}^{\text{dir}} / P_{\text{second}}^{\text{dir}}\right)\).  
(b) The ratio \(P_{\text{peak}}^{\text{MPS}} / P_{\text{second}}^{\text{MPS}}\) as a function of \(\chi\).  
The dashed line indicates the threshold value above which the hidden string can be easily identified.}\label{Rel_prob_1}
\end{figure}

Here, we investigate whether the MPS method can be effectively used to analyze the peaked circuits generated by the proposed method and to identify the hidden string. Specifically, we estimate the bond dimension \(\chi\) required to correctly find the hidden string as a function of the number of qubits \(n_\text{q}\) and the circuit depth \(n_\text{l}\). The key question is whether \(\chi\) can be chosen significantly smaller than the maximum value \(2^{n_\text{q}/2}\) while still reliably recovering the hidden string.

To analyze this, we introduce several parameters. We denote the probability of observing the hidden string in a direct simulation of the quantum circuit (without gate or measurement errors) as \(P_{\text{peak}}^{\text{dir}}\). The corresponding probability obtained after MPS simulation is \(P_{\text{peak}}^{\text{MPS}}\). Similarly, we define \(P_{\text{second}}^{\text{dir}}\) and \(P_{\text{second}}^{\text{MPS}}\) as the probabilities of observing the second most probable string in the direct and MPS simulations, respectively.

To obtain these parameters, we generate multiple quasi-random peaked circuits using the described method, varying \(n_\text{q}\) and \(n_\text{l}\). Simulations are performed without gate or measurement errors and without qubit decoherence. For each circuit, we perform multi-shot sampling with approximately \(10^5\) shots using either direct simulation or the MPS method. The string observed most frequently is identified as the hidden string, and its relative frequency defines \(P_{\text{peak}}\). The second most frequently observed string defines \(P_{\text{second}}\).

\begin{figure}[h]
\centering
\includegraphics[width=1\textwidth]{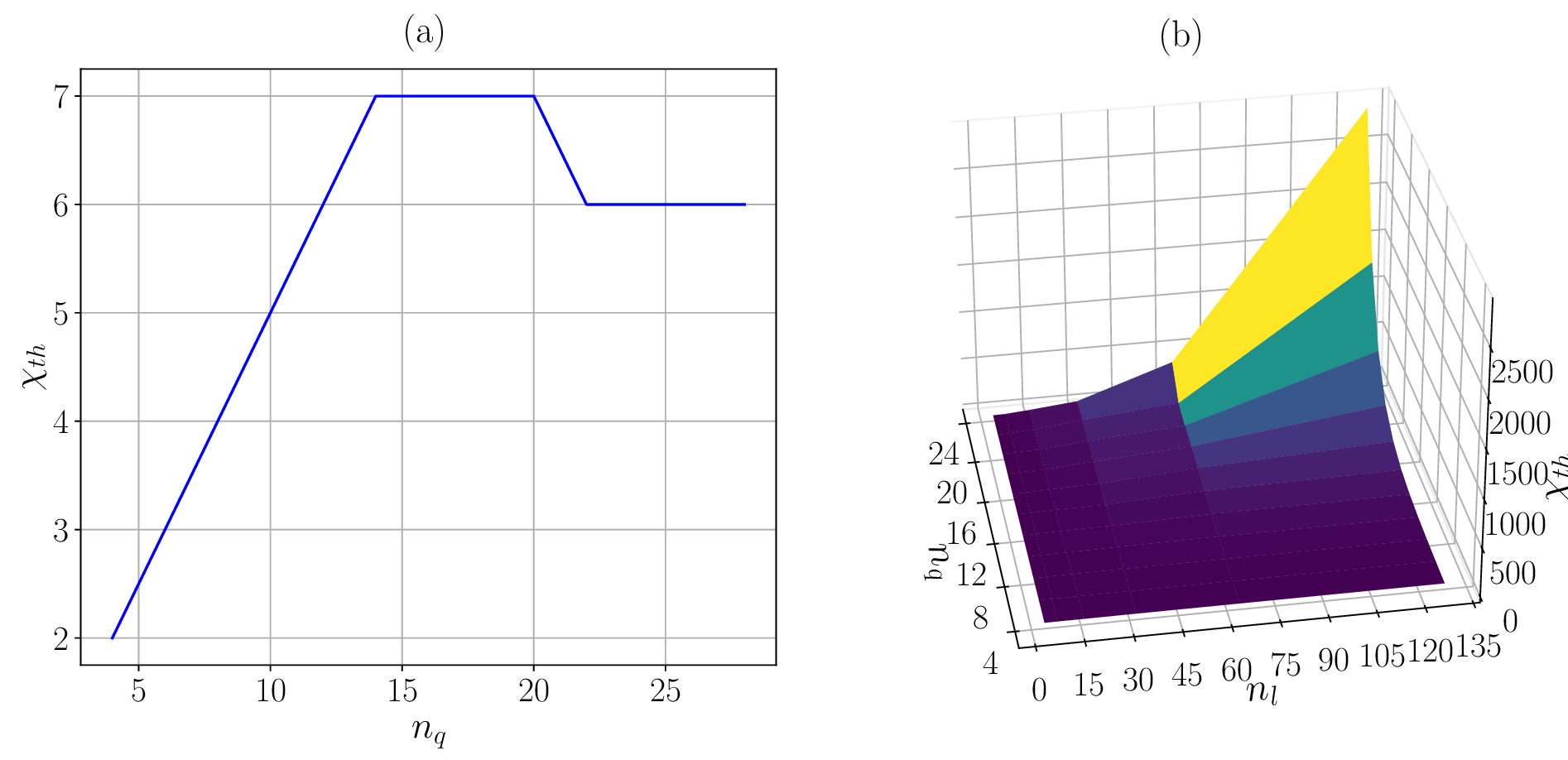}
\caption{Threshold value \(\chi_{\text{th}}\) of the bond dimension at which the hidden string can be identified using the MPS method.  
(a) \(\chi_{\text{th}}\) as a function of the number of qubits \(n_\text{q}\) for a fixed circuit depth \(n_\text{l} = 8\).  
(b) \(\chi_{\text{th}}\) as a function of both \(n_\text{q}\) and \(n_\text{l}\).}\label{fig_chi}
\end{figure}

Figure~\ref{Rel_prob_1}(a) shows how these parameters vary with the bond dimension \(\chi\). The solid lines represent the ratio \(P_{\text{peak}}^{\text{MPS}} / P_{\text{peak}}^{\text{dir}}\) for a fixed number of qubits (\(n_\text{q} = 8\)) but different circuit depths \(n_\text{l}\). The maximum possible \(\chi\) value for \(n_\text{q} = 8\) is 16, so at \(\chi = 16\) the MPS simulation should reproduce the results of the direct simulation, causing all curves to approach 1 at this point. However, for \(\chi < 16\), the ratio decreases more significantly for circuits with greater depth. 

The dashed lines depict the ratio \(\left(P_{\text{peak}}^{\text{MPS}} / P_{\text{second}}^{\text{MPS}}\right) / \left(P_{\text{peak}}^{\text{dir}} / P_{\text{second}}^{\text{dir}}\right)\). For some values of \(\chi\), this ratio reaches zero, indicating that the most probable string identified by the MPS simulation does not correspond to the hidden string.

The ratio \(P_{\text{peak}}^{\text{MPS}} / P_{\text{second}}^{\text{MPS}}\) as a function of \(\chi\) is shown in Fig.~\ref{Rel_prob_1}(b). We consider the circuit to be peaked if the probability of observing the correct hidden string is at least ten times higher than that of the second most probable string. The horizontal line denotes this threshold level. Even for a fixed number of qubits, the circuit depth significantly influences the minimum \(\chi\) value required to correctly identify the hidden string. Specifically, the longer the circuit, the closer the required \(\chi\) approaches its maximum possible value.

Figure~\ref{fig_chi}(a) shows how the minimum bond dimension \(\chi_\text{th}\) required to identify the hidden string depends on the number of qubits \(n_\text{q}\) for a fixed circuit depth, with an example depth of \(n_\text{l} = 8\). It can be observed that at low qubit numbers, the minimum \(\chi\) increases with \(n_\text{q}\), but then saturates as \(n_\text{q}\) grows further. This behavior aligns with recent findings on classical simulations of shallow quantum peaked circuits. Reference~\cite{Shallow_circuit_sim} demonstrated that circuits with a large number of qubits can be efficiently simulated classically if the circuits are shallow. Intuitively, this is because, in the shallow circuits, each qubit interacts (and is entangled) only with a limited number of nearest neighbors, and the number of these neighbors is constrained by the circuit depth.

Figure~\ref{fig_chi}(b) shows how the minimum bond dimension \(\chi_{\text{th}}\) depends on both the number of qubits \(n_\text{q}\) and the circuit depth \(n_\text{l}\). Again, saturation of \(\chi_{\text{th}}\) is observed for shallow circuits (low \(n_l\)), where increasing the number of qubits does not significantly increase the required \(\chi\). In this regime, the MPS method remains highly efficient for simulating random circuits. However, as the circuit depth increases, \(\chi_{\text{th}}\) approaches the maximum possible value \(2^{n_\text{q}/2}\). Thus, for deep circuits generated by the proposed method, the MPS approach is not expected to offer a significant efficiency advantage over direct simulation.

\section{Conclusion}\label{sec7}
In this work, we propose a method for generating quasi-random brick-wall quantum circuits that produce peaked output states - i.e., quantum circuits whose final states are highly concentrated on a specific computational basis state. This method enables the construction of peaked circuits with sizes that significantly exceed the limits of classical simulability. We also introduce a technique for concealing the target bitstring within the circuit, as well as a method for generating double-peaked circuits. The degree of peakedness - defined as the probability of measuring the hidden bitstring - can be tuned via a parameter in the algorithm. We demonstrate that the matrix product state (MPS) simulation method is generally ineffective at resolving the hidden bitstring in deep peaked circuits generated by our approach. However, MPS can be successfully applied to shallow peaked circuits.

\backmatter

\bmhead{Acknowledgements}

We would like to thank the BlueQubit team for organizing hackathons that inspired this study. Special thanks to Sergey Anishenko for insightful discussions and valuable input.

\section*{Declarations}

\begin{itemize}
\item Funding

Not applicable

\item Conflict of interest/Competing interests 

On behalf of all authors, the corresponding author states that there is no conflict of
interest.


\item Data availability

No datasets were generated or analyzed during the current study.

\item Materials availability

Not applicable

\item Code availability 

https://github.com/OlegUdalov/QC-qiskit-codes/tree/main/PeakedCircuits

\item Author contribution

All authors are contributed equally.

\end{itemize}







\bibliography{peaked-bibl}


\begin{thebibliography}{25}
\ifx \bisbn   \undefined \def \bisbn  #1{ISBN #1}\fi
\ifx \binits  \undefined \def \binits#1{#1}\fi
\ifx \bauthor  \undefined \def \bauthor#1{#1}\fi
\ifx \batitle  \undefined \def \batitle#1{#1}\fi
\ifx \bjtitle  \undefined \def \bjtitle#1{#1}\fi
\ifx \bvolume  \undefined \def \bvolume#1{\textbf{#1}}\fi
\ifx \byear  \undefined \def \byear#1{#1}\fi
\ifx \bissue  \undefined \def \bissue#1{#1}\fi
\ifx \bfpage  \undefined \def \bfpage#1{#1}\fi
\ifx \blpage  \undefined \def \blpage #1{#1}\fi
\ifx \burl  \undefined \def \burl#1{\textsf{#1}}\fi
\ifx \doiurl  \undefined \def \doiurl#1{\url{https://doi.org/#1}}\fi
\ifx \betal  \undefined \def \betal{\textit{et al.}}\fi
\ifx \binstitute  \undefined \def \binstitute#1{#1}\fi
\ifx \binstitutionaled  \undefined \def \binstitutionaled#1{#1}\fi
\ifx \bctitle  \undefined \def \bctitle#1{#1}\fi
\ifx \beditor  \undefined \def \beditor#1{#1}\fi
\ifx \bpublisher  \undefined \def \bpublisher#1{#1}\fi
\ifx \bbtitle  \undefined \def \bbtitle#1{#1}\fi
\ifx \bedition  \undefined \def \bedition#1{#1}\fi
\ifx \bseriesno  \undefined \def \bseriesno#1{#1}\fi
\ifx \blocation  \undefined \def \blocation#1{#1}\fi
\ifx \bsertitle  \undefined \def \bsertitle#1{#1}\fi
\ifx \bsnm \undefined \def \bsnm#1{#1}\fi
\ifx \bsuffix \undefined \def \bsuffix#1{#1}\fi
\ifx \bparticle \undefined \def \bparticle#1{#1}\fi
\ifx \barticle \undefined \def \barticle#1{#1}\fi
\bibcommenthead
\ifx \bconfdate \undefined \def \bconfdate #1{#1}\fi
\ifx \botherref \undefined \def \botherref #1{#1}\fi
\ifx \url \undefined \def \url#1{\textsf{#1}}\fi
\ifx \bchapter \undefined \def \bchapter#1{#1}\fi
\ifx \bbook \undefined \def \bbook#1{#1}\fi
\ifx \bcomment \undefined \def \bcomment#1{#1}\fi
\ifx \oauthor \undefined \def \oauthor#1{#1}\fi
\ifx \citeauthoryear \undefined \def \citeauthoryear#1{#1}\fi
\ifx \endbibitem  \undefined \def \endbibitem {}\fi
\ifx \bconflocation  \undefined \def \bconflocation#1{#1}\fi
\ifx \arxivurl  \undefined \def \arxivurl#1{\textsf{#1}}\fi
\csname PreBibitemsHook\endcsname

\bibitem[\protect\citeauthoryear{Arute et~al.}{2019}]{Intro_suprem_1}
\begin{barticle}
\bauthor{\bsnm{Arute}, \binits{F.}},
\bauthor{\bsnm{Arya}, \binits{K.}},
\bauthor{\bsnm{Babbush}, \binits{R.}},
\bauthor{\bsnm{Bacon}, \binits{D.}},
\bauthor{\bsnm{Bardin}, \binits{J.}},
\bauthor{\bsnm{Barends}, \binits{R.}},
\bauthor{\bsnm{Biswas}, \binits{R.}},
\bauthor{\bsnm{Boixo}, \binits{S.}},
\bauthor{\bsnm{Brandao}, \binits{F.}},
\bauthor{\bsnm{Buell}, \binits{D.}},
\bauthor{\bsnm{Burkett}, \binits{B.}},
\bauthor{\bsnm{Chen}, \binits{Y.}},
\bauthor{\bsnm{Chen}, \binits{Z.}},
\bauthor{\bsnm{Chiaro}, \binits{B.}},
\bauthor{\bsnm{Collins}, \binits{R.}},
\bauthor{\bsnm{Courtney}, \binits{W.}},
\bauthor{\bsnm{Dunsworth}, \binits{A.}},
\bauthor{\bsnm{Farhi}, \binits{E.}},
\bauthor{\bsnm{Foxen}, \binits{B.}},
\bauthor{\bsnm{Martinis}, \binits{J.}}:
\batitle{Quantum supremacy using a programmable superconducting processor}.
\bjtitle{Nature}
\bvolume{574},
\bfpage{505}--\blpage{510}
(\byear{2019})
\doiurl{10.1038/s41586-019-1666-5}
\end{barticle}
\endbibitem

\bibitem[\protect\citeauthoryear{Aaronson and
  Chen~Geddes}{2017}]{Intro_suprem_2}
\begin{barticle}
\bauthor{\bsnm{Aaronson}, \binits{S.}},
\bauthor{\bsnm{Chen~Geddes}, \binits{L.}}:
\batitle{Complexity-theoretic foundations of quantum supremacy experiments}.
\bjtitle{Computational Complexity Conference (CCC 2017), ser. LIPIcs}
\bvolume{79},
\bfpage{1}--\blpage{67}
(\byear{2017})
\end{barticle}
\endbibitem

\bibitem[\protect\citeauthoryear{Movassagh}{2019}]{Intro_suprem_3}
\begin{botherref}
\oauthor{\bsnm{Movassagh}, \binits{R.}}:
Quantum supremacy and random circuits.
arXiv
(1909.06210)
(2019)
\end{botherref}
\endbibitem

\bibitem[\protect\citeauthoryear{Preskill}{2012}]{Intro_suprem_4}
\begin{botherref}
\oauthor{\bsnm{Preskill}, \binits{J.}}:
Quantum computing and the entanglement frontier.
arXiv
(1203.5813)
(2012)
\end{botherref}
\endbibitem

\bibitem[\protect\citeauthoryear{Bouland et~al.}{2018}]{Intro_suprem_5}
\begin{botherref}
\oauthor{\bsnm{Bouland}, \binits{A.}},
\oauthor{\bsnm{Fefferman}, \binits{B.}},
\oauthor{\bsnm{Nirkhe}, \binits{C.}},
\oauthor{\bsnm{U.}, \binits{V.}}:
Quantum supremacy and the complexity of random circuit sampling.
arXiv
(1803.04402)
(2018)
\end{botherref}
\endbibitem

\bibitem[\protect\citeauthoryear{Shor}{1994}]{Shor}
\begin{bchapter}
\bauthor{\bsnm{Shor}, \binits{P.W.}}:
\bctitle{Algorithms for quantum computation: discrete logarithms and
  factoring}.
In: \bbtitle{Proceedings 35th Annual Symposium on Foundations of Computer
  Science},
pp. \bfpage{124}--\blpage{134}
(\byear{1994}).
\doiurl{10.1109/SFCS.1994.365700}
\end{bchapter}
\endbibitem

\bibitem[\protect\citeauthoryear{Aaronson and Zhang}{2024}]{Peaked_main}
\begin{botherref}
\oauthor{\bsnm{Aaronson}, \binits{S.}},
\oauthor{\bsnm{Zhang}, \binits{Y.}}:
On verifiable quantum advantage with peaked circuit sampling.
arXiv
(2404.14493)
(2024)
\end{botherref}
\endbibitem

\bibitem[\protect\citeauthoryear{Liu and Wang}{2018}]{PhysRevA.98.062324}
\begin{barticle}
\bauthor{\bsnm{Liu}, \binits{J.-G.}},
\bauthor{\bsnm{Wang}, \binits{L.}}:
\batitle{Differentiable learning of quantum circuit born machines}.
\bjtitle{Phys. Rev. A}
\bvolume{98},
\bfpage{062324}
(\byear{2018})
\doiurl{10.1103/PhysRevA.98.062324}
\end{barticle}
\endbibitem

\bibitem[\protect\citeauthoryear{Benedetti
  et~al.}{2019}]{Benedetti2019Generative}
\begin{barticle}
\bauthor{\bsnm{Benedetti}, \binits{M.}},
\bauthor{\bsnm{Garcia-Pintos}, \binits{D.}},
\bauthor{\bsnm{Perdomo}, \binits{O.}},
\bauthor{\bsnm{Leyton-Ortega}, \binits{V.}},
\bauthor{\bsnm{Nam}, \binits{Y.}},
\bauthor{\bsnm{Perdomo-Ortiz}, \binits{A.}}:
\batitle{A generative modeling approach for benchmarking and training shallow
  quantum circuits}.
\bjtitle{npj Quantum Information}
\bvolume{5},
\bfpage{45}
(\byear{2019})
\doiurl{10.1038/s41534-019-0157-8}
\end{barticle}
\endbibitem

\bibitem[\protect\citeauthoryear{Du et~al.}{2022}]{Du2022Power}
\begin{botherref}
\oauthor{\bsnm{Du}, \binits{Y.}},
\oauthor{\bsnm{Tu}, \binits{Z.}},
\oauthor{\bsnm{Wu}, \binits{B.}},
\oauthor{\bsnm{Yuan}, \binits{X.}},
\oauthor{\bsnm{Tao}, \binits{D.}}:
Power of quantum generative learning.
arXiv preprint arXiv:2205.04730
(2022)
\end{botherref}
\endbibitem

\bibitem[\protect\citeauthoryear{Coyle et~al.}{2021}]{Coyle2021Quantum}
\begin{barticle}
\bauthor{\bsnm{Coyle}, \binits{B.}},
\bauthor{\bsnm{Henderson}, \binits{M.}},
\bauthor{\bsnm{Le}, \binits{J.C.J.}},
\bauthor{\bsnm{Kumar}, \binits{N.}},
\bauthor{\bsnm{Paini}, \binits{M.}},
\bauthor{\bsnm{Kashefi}, \binits{E.}}:
\batitle{Quantum versus classical generative modelling in finance}.
\bjtitle{Quantum Science and Technology}
\bvolume{6}(\bissue{2}),
\bfpage{024013}
(\byear{2021})
\doiurl{10.1088/2058-9565/abd2db}
\end{barticle}
\endbibitem

\bibitem[\protect\citeauthoryear{Wecker et~al.}{2016}]{Wecker2016Training}
\begin{barticle}
\bauthor{\bsnm{Wecker}, \binits{D.}},
\bauthor{\bsnm{Hastings}, \binits{M.B.}},
\bauthor{\bsnm{Troyer}, \binits{M.}}:
\batitle{Training a quantum optimizer}.
\bjtitle{Physical Review A}
\bvolume{94}(\bissue{2}),
\bfpage{022309}
(\byear{2016})
\doiurl{10.1103/PhysRevA.94.022309}
\end{barticle}
\endbibitem

\bibitem[\protect\citeauthoryear{Gharibyan
  et~al.}{2023}]{Gharibyan2023Hierarchical}
\begin{barticle}
\bauthor{\bsnm{Gharibyan}, \binits{H.}},
\bauthor{\bsnm{Su}, \binits{V.}},
\bauthor{\bsnm{Tepanyan}, \binits{H.}}:
\batitle{Hierarchical learning for quantum ml: Novel training technique for
  large-scale variational quantum circuits}.
\bjtitle{arXiv preprint arXiv:2311.12929}
(\byear{2023})
\doiurl{10.48550/arXiv.2311.12929}
\end{barticle}
\endbibitem

\bibitem[\protect\citeauthoryear{Berthusen
  et~al.}{2022}]{PhysRevResearch.4.023097}
\begin{barticle}
\bauthor{\bsnm{Berthusen}, \binits{N.F.}},
\bauthor{\bsnm{Trevisan}, \binits{T.V.}},
\bauthor{\bsnm{Iadecola}, \binits{T.}},
\bauthor{\bsnm{Orth}, \binits{P.P.}}:
\batitle{Quantum dynamics simulations beyond the coherence time on noisy
  intermediate-scale quantum hardware by variational trotter compression}.
\bjtitle{Phys. Rev. Res.}
\bvolume{4},
\bfpage{023097}
(\byear{2022})
\doiurl{10.1103/PhysRevResearch.4.023097}
\end{barticle}
\endbibitem

\bibitem[\protect\citeauthoryear{Miháliková
  et~al.}{2025}]{Mihalikova2025Impact}
\begin{barticle}
\bauthor{\bsnm{Miháliková}, \binits{I.}},
\bauthor{\bsnm{Krejčí}, \binits{M.}},
\bauthor{\bsnm{Friák}, \binits{M.}}:
\batitle{The impact of quantum circuit architecture and hyperparameters on
  variational quantum algorithms exemplified in the electronic structure of the
  gaas crystal}.
\bjtitle{Scientific Reports}
\bvolume{15},
\bfpage{15746}
(\byear{2025})
\doiurl{10.1038/s41598-025-00151-x}
\end{barticle}
\endbibitem

\bibitem[\protect\citeauthoryear{Uvarov et~al.}{2020}]{PhysRevA.102.012415}
\begin{barticle}
\bauthor{\bsnm{Uvarov}, \binits{A.V.}},
\bauthor{\bsnm{Kardashin}, \binits{A.S.}},
\bauthor{\bsnm{Biamonte}, \binits{J.D.}}:
\batitle{Machine learning phase transitions with a quantum processor}.
\bjtitle{Phys. Rev. A}
\bvolume{102},
\bfpage{012415}
(\byear{2020})
\doiurl{10.1103/PhysRevA.102.012415}
\end{barticle}
\endbibitem

\bibitem[\protect\citeauthoryear{Emerson et~al.}{2005}]{Emerson2005Scalable}
\begin{barticle}
\bauthor{\bsnm{Emerson}, \binits{J.}},
\bauthor{\bsnm{Alicki}, \binits{R.}},
\bauthor{\bsnm{Życzkowski}, \binits{K.}}:
\batitle{Scalable noise estimation with random unitary operators}.
\bjtitle{Journal of Optics B: Quantum and Semiclassical Optics}
\bvolume{7}(\bissue{10}),
\bfpage{347}--\blpage{352}
(\byear{2005})
\doiurl{10.1088/1464-4266/7/10/021}
\end{barticle}
\endbibitem

\bibitem[\protect\citeauthoryear{Elben et~al.}{2023}]{Elben2023Randomized}
\begin{barticle}
\bauthor{\bsnm{Elben}, \binits{A.}},
\bauthor{\bsnm{Flammia}, \binits{S.T.}},
\bauthor{\bsnm{Huang}, \binits{H.-Y.}},
\bauthor{\bsnm{Kueng}, \binits{R.}},
\bauthor{\bsnm{Preskill}, \binits{J.}},
\bauthor{\bsnm{Vermersch}, \binits{B.}},
\bauthor{\bsnm{Zoller}, \binits{P.}}:
\batitle{The randomized measurement toolbox}.
\bjtitle{Nature Reviews Physics}
\bvolume{5},
\bfpage{9}--\blpage{24}
(\byear{2023})
\doiurl{10.1038/s42254-022-00535-2}
\end{barticle}
\endbibitem

\bibitem[\protect\citeauthoryear{Javadi-Abhari
  et~al.}{2024}]{JavadiAbhari2024Qiskit}
\begin{barticle}
\bauthor{\bsnm{Javadi-Abhari}, \binits{A.}},
\bauthor{\bsnm{Treinish}, \binits{M.}},
\bauthor{\bsnm{Krsulich}, \binits{K.}},
\bauthor{\bsnm{Wood}, \binits{C.J.}},
\bauthor{\bsnm{Lishman}, \binits{J.}},
\bauthor{\bsnm{Gacon}, \binits{J.}},
\bauthor{\bsnm{Martiel}, \binits{S.}},
\bauthor{\bsnm{Nation}, \binits{P.D.}},
\bauthor{\bsnm{Bishop}, \binits{L.S.}},
\bauthor{\bsnm{Cross}, \binits{A.W.}},
\bauthor{\bsnm{Johnson}, \binits{B.R.}},
\bauthor{\bsnm{Gambetta}, \binits{J.M.}}:
\batitle{Quantum computing with qiskit}.
\bjtitle{arXiv preprint arXiv:2405.08810}
(\byear{2024})
\doiurl{10.48550/arXiv.2405.08810}
{\href{https://arxiv.org/abs/2405.08810}{{arXiv:2405.08810}}}
\end{barticle}
\endbibitem

\bibitem[\protect\citeauthoryear{Xu et~al.}{2023}]{Xu2023Herculean}
\begin{barticle}
\bauthor{\bsnm{Xu}, \binits{X.}},
\bauthor{\bsnm{Benjamin}, \binits{S.}},
\bauthor{\bsnm{Sun}, \binits{J.}},
\bauthor{\bsnm{Yuan}, \binits{X.}},
\bauthor{\bsnm{Zhang}, \binits{P.}}:
\batitle{A herculean task: Classical simulation of quantum computers}.
\bjtitle{arXiv preprint arXiv:2302.08880}
(\byear{2023})
\doiurl{10.48550/arXiv.2302.08880}
{\href{https://arxiv.org/abs/2302.08880}{{arXiv:2302.08880}}}
\end{barticle}
\endbibitem

\bibitem[\protect\citeauthoryear{Ba\~nuls et~al.}{2006}]{PhysRevA.73.022344}
\begin{barticle}
\bauthor{\bsnm{Ba\~nuls}, \binits{M.C.}},
\bauthor{\bsnm{Or\'us}, \binits{R.}},
\bauthor{\bsnm{Latorre}, \binits{J.I.}},
\bauthor{\bsnm{P\'erez}, \binits{A.}},
\bauthor{\bsnm{Ruiz-Femen\'{\i}a}, \binits{P.}}:
\batitle{Simulation of many-qubit quantum computation with matrix product
  states}.
\bjtitle{Phys. Rev. A}
\bvolume{73},
\bfpage{022344}
(\byear{2006})
\doiurl{10.1103/PhysRevA.73.022344}
\end{barticle}
\endbibitem

\bibitem[\protect\citeauthoryear{Perez-Garcia
  et~al.}{2007}]{PerezGarcia2007MPS}
\begin{barticle}
\bauthor{\bsnm{Perez-Garcia}, \binits{D.}},
\bauthor{\bsnm{Verstraete}, \binits{F.}},
\bauthor{\bsnm{Wolf}, \binits{M.M.}},
\bauthor{\bsnm{Cirac}, \binits{J.I.}}:
\batitle{Matrix product state representations}.
\bjtitle{ArXiv}
(\byear{2007})
\doiurl{10.48550/arXiv.quant-ph/0608197}
{[quant-ph]}
\end{barticle}
\endbibitem

\bibitem[\protect\citeauthoryear{Martin et~al.}{2024}]{Martin2024}
\begin{barticle}
\bauthor{\bsnm{Martin}, \binits{A.}},
\bauthor{\bsnm{Ayral}, \binits{T.}},
\bauthor{\bsnm{Jamet}, \binits{F.}},
\bauthor{\bsnm{Rančić}, \binits{M.J.}},
\bauthor{\bsnm{Simon}, \binits{P.}}:
\batitle{Combining matrix product states and noisy quantum computers for
  quantum simulation}.
\bjtitle{Physical Review A}
\bvolume{109}(\bissue{6}),
\bfpage{062437}
(\byear{2024})
\doiurl{10.1103/PhysRevA.109.062437}
{\href{https://arxiv.org/abs/2305.19231}{{arXiv:2305.19231}}}
{[quant‑ph]}
\end{barticle}
\endbibitem

\bibitem[\protect\citeauthoryear{Patra et~al.}{2024}]{PhysRevResearch.6.013326}
\begin{barticle}
\bauthor{\bsnm{Patra}, \binits{S.}},
\bauthor{\bsnm{Jahromi}, \binits{S.S.}},
\bauthor{\bsnm{Singh}, \binits{S.}},
\bauthor{\bsnm{Orus}, \binits{R.}}:
\batitle{Efficient tensor network simulation of ibm's largest quantum
  processors}.
\bjtitle{Phys. Rev. Res.}
\bvolume{6},
\bfpage{013326}
(\byear{2024})
\doiurl{10.1103/PhysRevResearch.6.013326}
\end{barticle}
\endbibitem

\bibitem[\protect\citeauthoryear{Bravyi et~al.}{2024}]{Shallow_circuit_sim}
\begin{bchapter}
\bauthor{\bsnm{Bravyi}, \binits{S.}},
\bauthor{\bsnm{Gosset}, \binits{D.}},
\bauthor{\bsnm{Liu}, \binits{Y.}}:
\bctitle{Classical simulation of peaked shallow quantum circuits}.
In: \bbtitle{Proceedings of the 56th Annual ACM Symposium on Theory of
  Computing}.
\bsertitle{STOC 2024},
pp. \bfpage{561}--\blpage{572}.
\bpublisher{Association for Computing Machinery},
\blocation{New York, NY, USA}
(\byear{2024}).
\doiurl{10.1145/3618260.3649638} .
\burl{https://doi.org/10.1145/3618260.3649638}
\end{bchapter}
\endbibitem

\end{thebibliography}

\end{document}